%% file: main.tex
\documentclass[runningheads]{llncs}
\pdfoutput=1
\usepackage{graphicx}
\usepackage{microtype}
\usepackage{subfigure}
\usepackage{booktabs}
\usepackage{mathtools}
\usepackage[utf8]{inputenc} 
\usepackage[T1]{fontenc}
\usepackage{url}
\usepackage{booktabs}    
\usepackage{amsfonts}     
\usepackage{nicefrac}       
\usepackage{microtype}      
\usepackage{xcolor}       
\usepackage[pagebackref,breaklinks,colorlinks,citecolor=blue]{hyperref}
\usepackage{amsmath}
\usepackage{multirow}
\usepackage{multicol}
\usepackage{amssymb}
\usepackage{longtable}
\usepackage{graphicx}
\usepackage{bbding}
\usepackage{soul}
\usepackage{enumitem}
\usepackage[most]{tcolorbox}
\usepackage[ruled,linesnumbered]{algorithm2e}

\setitemize[1]{itemsep=5pt,partopsep=0pt,parsep=\parskip,topsep=5pt}
\setenumerate[1]{itemsep=5pt,partopsep=0pt,parsep=\parskip,topsep=5pt}

\begin{document}

\title{MILE: A Mutation Testing Framework of In-Context Learning Systems}

\author{Zeming Wei, Yihao Zhang, and Meng Sun\inst{(}\Envelope\inst{)}}
\authorrunning{Z. Wei et al.}
%
\institute{Peking University, Beijing 100871, China\\
\email{weizeming@stu.pku.edu.cn},\\
\email {zhangyihao@stu.pku.edu.cn},\\
\email{sunmeng@math.pku.edu.cn}}
\maketitle
\begin{abstract}
In-context Learning (ICL) has achieved notable success in the applications of large language models (LLMs). By adding only a few input-output pairs that demonstrate a new task, the LLM can efficiently learn the task during inference without modifying the model parameters. Such mysterious ability of LLMs has attracted great research interests in understanding, formatting, and improving the in-context demonstrations, while still suffering from drawbacks like black-box mechanisms and sensitivity against the selection of examples. In this work, inspired by the foundations of adopting testing techniques in machine learning (ML) systems, we propose a mutation testing framework designed to characterize the quality and effectiveness of test data for ICL systems. First, we propose several mutation operators specialized for ICL demonstrations, as well as corresponding mutation scores for ICL test sets. With comprehensive experiments, we showcase the effectiveness of our framework in evaluating the reliability and quality of ICL test suites. Our code is available at \url{https://github.com/weizeming/MILE}.

\end{abstract}

\input{1_intro}

\input{2_pre}

\input{3_pipeline}

\input{4_experiments}

\input{5_related}

\input{6_conclusion}

\section*{Acknowlegement}
This work was sponsored by the National Natural Science Foundation of China (Grant No. 62172019) and the Beijing Natural Science Foundation's Undergraduate Initiating Research Program (Grant No. QY23041).

\newpage

\bibliographystyle{splncs04}
\bibliography{reference.bib}
\end{document}

%% file: 1_intro.tex
\section{Introduction}

In the past few years, Large Language Models (LLMs)~\cite{achiam2023gpt,zheng2024judging,touvron2023llama,almazrouei2023falcon} have achieved milestone success across a variety of tasks~\cite{huang2024understanding,wu2023autogen,wangexecutable}. In particular, the In-Context Learning (ICL)~\cite{brown2020language,dong2023survey} property of LLMs has been recognized as a key emerging ability of LLMs~\cite{lu2023emergent,schaeffer2024emergent}. By prompting a few input-label demonstrations as the context, LLMs can be adapted efficiently to new tasks \emph{without} modifying any model parameters. This enigmatic characteristic of LLMs has sparked significant research interest in comprehending~\cite{xie2021explanation,min2022rethinking,dai2023can,wang2023large,garg2022can} and utilizing~\cite{wei2022chain,wei2023jailbreak,pawelczyk2023context,wang2024theoretical} ICL in diverse scenarios.

However, ICL has been shown to have notable reliability issues, such as strong dependence on the selection of examples~\cite{agrawal2022context}, the order sensitivity~\cite{lu2021fantastically,zhao2021calibrate} of the demonstrations, and vulnerabilities against adversarial attacks~\cite{wei2023jailbreak,wang2023adversarial,qiang2023hijacking}.To mitigate these issues, a series of works have been proposed to automatically organize demonstrations~\cite{lu2021fantastically,agrawal2022context} or design intrinsically robust ICL mechanisms~\cite{ratner2023parallel,zhang2024batch,fang2024rethinking}. While these works mainly focus on improving the robustness of ICL, how to select high-quality test suites for evaluating ICL systems remains an open research problem. Moreover, as the computational cost of LLMs becomes significantly higher than that of conventional deep neural networks~\cite{yang2024harnessing}, the need for high-quality datasets to conduct more efficient and accurate evaluations is emphasized even further.

On the other hand, mutation testing~\cite{jia2010analysis} techniques have showcased impressive potential in studying the reliability defects and test suite quality of machine learning (ML) systems~\cite{ma2018deepmutation,wang2019adversarial,zhang2020machine,huang2020survey}. By regarding the ML system as the software under test (SUT)~\cite{demillo1991constraint}, several mutation testing methods have been designed for different ML paradigms including deep learning~\cite{ma2018deepmutation,shen2018munn,hu2029deep}, reinforcement learning~\cite{uesato2018rigorous,lu2022towards}, and unsupervised learning~\cite{lu2024mutation}. Specifically, similar to mutation testing for general software systems, these methods apply mutators particularly designed for the machine learning models or training data, and then study the behavior differences between the original model and
the mutant models. Since the primary goal of mutation testing is to assess the efficacy of test cases in characterizing faults in the ML model, test suites showcasing superior performance disparities between the original model and the mutant models are deemed of better quality.

In this paper, driven by the observation that ICL systems also encounter robustness issues and demand high-quality test cases, we propose \textbf{MILE}, a \textbf{M}utation testing framework for \textbf{I}n-context \textbf{LE}arning systems. First, we propose mutators specialized for ICL systems. Unlike mutation testing on conventional deep learning systems that consider both data and model mutators~\cite{ma2018deepmutation,hu2029deep}, we primarily focus on mutation operations on the ICL prompt since ICL systems typically use a static pre-trained LLM and mainly concentrate on designing demonstrations. Taking into account the characteristics of ICL, such as sensitivity to the orders and strong dependence on the labels, we propose a kit of mutators including demonstration-level ones and prompt-level ones. Meanwhile, we design corresponding mutation scores for MILE. Besides classic mutation scores, we propose a group-wise mutation score that takes into consideration the diversity of defects within the prompt. This score is helpful for identifying how well test suites can characterize diverse defects, beyond just evaluating the test set as a whole.

We finally evaluate our MILE framework across benchmark datasets and popular LLMs. Similar to existing mutation testing frameworks~\cite{ma2018deepmutation,hu2029deep}, we sample test data from uniform or non-uniform classes to simulate high- or low-quality datasets and calculate the mutation scores on them. The experiment results suggest that our mutation scores have a strong correlation to the quality of the test sets, showcasing the effectiveness of our framework in measuring the quality of test suites. In addition, we take an in-depth analysis of each mutator to better understand their sensitivity to the defects within the ICL prompts, which is helpful for mutation operation selection and allocation for testing ICL systems with different scenarios.

Overall, our contributions in this work can be summarized as follows:

\begin{enumerate}
    \item We propose a mutation testing framework of in-context learning systems, named MILE, to comprehensively assess the effectiveness and quality of the test cases for ICL.
    \item We design demonstration-level and prompt-level mutation operators based on the characteristics of ICL. Furthermore, we propose standard and group-wise mutation scores for better evaluation.
    \item We implement MILE and evaluate it across benchmark datasets and LLMs to showcase its effectiveness in assessing the quality of test cases. We also conduct an analysis for each independent mutator.
\end{enumerate}

The rest of this paper is organized as follows. We start by briefly introducing the backgrounds and related notations for ICL and mutation testing in Section~\ref{sec:pre}. In Section~\ref{sec:framework}, we first provide an overview of our framework, followed by introducing our mutation operators and scores designed for ICL. We then present our evaluation of MILE in Section~\ref{sec:experiment}, including experiment set-up, overall assessment, and independent mutator analysis. Finally, we discuss related work in Section~\ref{sec:related} and conclude our work in Section~\ref{sec:conclusion}.

%% file: 2_pre.tex
\section{Preliminaries}
\label{sec:pre}
In this section, we provide background information and define formal notations for ICL and mutation testing.

\subsubsection{In-context learning (ICL).}
ICL~\cite{brown2020language,dong2023survey} is an intriguing property that emerges in LLMs in which they learn a specific task demonstrated by a few input-label pair examples. By keeping the model parameters static, prompting a system message that briefly describes the task and a set of input-label pairs demonstrating the task, the LLM can learn a mapping between the inputs and labels, and then successfully predict the label of a new input query attached behind the demonstrations in the prompt. Specifically, the definition of an ICL system can be formulated as follows:

\begin{definition}[In-context Learning System] An ICL system consists of a pre-trained LLM $M(\cdot)$ that returns a response $M(p)$ for any prompt $p$, a system prompt $p_s$, and a set of in-context demonstrations $D=\{(x_1, y_1), (x_2, y_2),\cdots, (x_k, y_k)\}$. For any test prompt $x_{test}$, the model gathers all sources to form the ICL prompt $p^*=[p_s, x_1, y_1, x_2, y_2, \cdots, x_k, y_k, x_{test}]$ and return the final response by $M(p^*)$.
\end{definition}

An example of an ICL prompt for the RTE task~\cite{dagan2005pascal} is illustrated in the following block. In this task, the goal is to determine whether the hypotheses can be derived from the premises, as instructed in the system message (lines 1-2). Then, 2 demonstrations consisting of inputs (the premises and hypotheses) and labels (answer ↑ or ↓) are attached behind the system prompt. Generally, the shots (number) of demonstrations are much more than 2. Finally, the prompt ends with a querying input for inference. From the text, we can know that the hypothesis \textit{``Qatar is located in Doha''} cannot be derived from the premise, which is the same as the \textit{2nd} demonstration, so the correct output from the model should be ↓.

\begin{center}
\begin{tcolorbox}[colback = black!10,width=\textwidth,title=Example ICL prompt for RTE task,left=5pt,right=0pt,top=0pt,bottom=0pt]
<s> Determine whether the hypotheses made based on the premises below are ↑ or ↓.\\
\textbf{Premise}: The Democrats' success in the 2006 elections means changes at the top in the House and Senate.\\
\textbf{Hypothesis}: Democrats won the 2006 elections.\\
\textbf{Answer}: ↑\\\\
\textbf{Premise}: IKEA offers fantastic and affordable solutions for your home furnishing needs.\\
\textbf{Hypothesis}: Ikea is a home.\\
\textbf{Answer}: ↓\\\\
\textbf{Premise}: VCU School of the Arts In Qatar is located in Doha, the capital city of Qatar.\\
\textbf{Hypothesis}: Qatar is located in Doha.\\
\textbf{Answer}:\vspace{5pt}
\end{tcolorbox}
\end{center}

\subsubsection{Mutation Testing.}
Test cases play a crucial role in characterizing and evaluating the vulnerability and reliability of software systems. As a pioneering technique, mutation testing was first proposed in the 1970s~\cite{offutt2001mutation,jia2010analysis,just2014defects4j} to measure the quality of test suites for software systems. Generally, mutation testing aims to replicate potential faults and vulnerabilities in the system to determine which test cases can effectively detect them. To this end, the mutation testing first artificially mutates a normal system to introduce fault with a set of pre-defined mutation operators (mutators). Then, given a test suite, its quality judged by this testing framework is determined by the ratio of the mutants that are killed by this dataset, as formally stated in the following definition.

\begin{definition}[Mutation Testing]
    Consider a program $P$, a set of mutation operators $O=\{o_1,o_2,\cdots,o_m\}$, and a test set $T=\{(X_1,Y_1), (X_2,Y_2),\cdots, (X_n,Y_n)\}$ where each $X_i$ is an input and each $Y_i$ is a label. With each mutator $o_i$ turns the program $P$ into a mutant program $o_i(P) = P'_i$, a mutation testing process evaluates $o_i(P)$ on all $(X_i,Y_i)$ and studies the difference between the performance of $P$ and the mutants $\{o_1(P), o_2(P), \cdots, o_m(P)\}$.
\end{definition}

So far, mutation testing has been acknowledged as one of the most fundamental software testing techniques, which is widely adopted 
in scenarios like fault localization~\cite{papadakis2015metallaxis} and software repairment~\cite{ghanbari2019practical}. In particular, mutation testing has proven to be successful in evaluating the adequacy of test datasets by providing a metric to determine whether existing tests have good fault-revealing capabilities. In the context of ML systems, a representative application of mutation testing is to assess the quality of test sets by treating the model as a program, and when the mutated models (mutants) output false prediction, this mutant can be regarded as \textit{killed}. We provide more related work on applying mutation testing for ML systems in Section~\ref{sec:related}.

%% file: 3_pipeline.tex
\section{Mutation Testing For In-Context Learning}
\label{sec:framework}

In this section, we present MILE, our mutation testing framework for in-context learning systems. We begin with a brief overview of the testing pipeline and general design for mutation operator and score, then put forward our solutions to them respectively.

\subsection{Overview}

Similar to existing mutation testing techniques for ML systems, we devise a two-stage testing framework consisting of mutant generation and test set evaluation. However, in contrast to traditional machine learning approaches that train models from scratch (\textit{i.e.} with random parameter initialization), ICL systems usually use a pre-trained static LLM and concentrate on creating in-context demonstrations. As a result, we only consider mutations in the demonstrations while keeping the LLM unchanged. 

The overall pipeline of our proposed MILE is elaborated in Algorithm~\ref{alg:overview}. 
In line 1, we first obtain the mutated in-context demonstrations $D'_i$ from $D$ with all mutators. Then, by incorporating these demonstrations into the original LLM $M$, we obtain ICL models $\mathcal M$ and $\mathcal M'_i$ (line 2).
\begin{algorithm}[t]
\SetKwInOut{Input}{Input}
\SetKwInOut{Output}{Output} 
\Input{LLM $M$, System prompt $p_s$, In-context demonstrations $D=\{(x_1, y_1), (x_2, y_2),\cdots, (x_k, y_k)\}$, Test set under evaluation $T=\{(X_1,Y_1), (X_2,Y_2)\cdots,(X_n,Y_n)\}$, Mutation operators $O=\{o_1,o_2,\cdots, o_m\}$}
\Output{Mutation scores and analysis}

Obtain mutant prompts $D'_i \gets o_i(D),\quad i=1,2,\cdots,m$\;

Construct ICL model $\mathcal M(\cdot) = M([p_s,D,\cdot])$ and mutant models $\mathcal M'_i(\cdot) = M([p_s,D'_i,\cdot]), \quad i=1,2,\cdots,m$\;

Mutant\_Outputs$\gets[\ ]$\;
\For{$(X_i,Y_i) \in T$}{
\If{$\mathcal M(X_i)=Y_i$}{
    Mutant\_Outputs.$append([\mathcal M'_1(X_1), \mathcal M'_2(X_i), \cdots, \mathcal M'_m(X_i)])$
}\Else{
\textbf{continue}\;
}
}
\textbf{return} Mutation\_Score(Mutant\_Outputs, $[Y_1,Y_2,\cdots,Y_n]$)\;
\caption{Pipeline of MLIE}
\label{alg:overview}
\end{algorithm}
The second stage is to evaluate the test set $T$ with the mutants. In line 5, we first filter out the examples that are misclassified by the original model. Following existing work~\cite{ma2018deepmutation}, we primarily illustrate our framework on classification tasks, but it can be easily adapted to other scenarios like regression tasks by adding a threshold function. Further, in line 6 for these passed test cases, we track all mutant predictions on them and finally calculate the mutation scores based on these outputs and true labels, as detailed in the following sections.

\subsection{Mutation Operators for ICL}

In this section, we propose several mutation operators specialized for ICL prompts. Considering the principle of the mutation operator, which is to characterize potential faults and the sensitivity of a program that may have suffered, we design mutators based on possible problems and the sensitivity of ICL prompts, and divide them into demonstration-level and prompt-level ones.

\subsubsection{Demonstration-level mutation operators.}

First, we consider demonstration-level mutations for a single demonstration $(x_i, y_i)$ that modify $x_i$ or $y_i$ to construct a mutant ICL prompt, including:
\begin{itemize}
    \item \textbf{Noisy Labels (NL)}. ICL is known to be sensitive to the noise of labels in the demonstrations~\cite{cheng2024exploring,gao2024noise}. However, recent research emphasizes the potential of scaling ICL to very large volumes~\cite{agarwal2024many,anil2024many} where ensuring label accuracy becomes challenging, leading to potential concerns about noisy labels within the prompt. Therefore, we first propose a Noisy Label (NL) mutator which randomly replaces a correct label in the demonstration: $y_i \gets y', i\sim Uniform([1...k]), y'\in\mathcal Y  - \{y_i\}$ where $\mathcal Y $ is all class labels.
    
    \item \textbf{Out-of-distribution Labels (OL)}. Similar to the Noisy Labels mutator, we also consider another common reliability issue that the label assigned to data may be out-of-distribution (OOD), as the OOD detection is still a not fully addressed problem~\cite{liu2020energy,yang2024generalized}. Unlike the NL mutator which injects a false label, this Out-of-distribution Labels mutator replaces the original label with one that does not belong to the task classes, \textit{e.g.} a special token: $y_i \gets z, i\sim Uniform([1...k]), z\not\in \mathcal Y $. Intuitively, the OOD label mutator may be more moderate than the noisy label mutator, as verified in our experiments.
    
    \item \textbf{Blurred Inputs (BI)}. In addition to mutating the labels in the demonstrations, we further consider potential issues in the inputs $x_i$. As stated in prior research~\cite{min2022rethinking}, high-quality inputs are essential for helping the language model better understand the task. Therefore, we suggest simulating questionable inputs in the demonstrations by blurring the input content: $x_i \gets \Tilde{x}_i, i\sim Uniform([1...k])$. In our implementation, we achieve this by simply truncating the input to its prefix.
\end{itemize}

\subsubsection{Prompt-level mutation operators.} We also consider prompt-level mutation, where we maintain input-label pairs for each individual demonstration but explore mutating between different demonstrations, including:
\begin{itemize}
    \item \textbf{Demonstration Shuffle (DS)}. The order of the demonstrations can have a significant impact on the ICL prompts, as noted in previous studies ~\cite{lu2021fantastically,fang2024rethinking}. Therefore, test cases for which the prediction changed after re-ordering the demonstrations would be considered as being near the decision boundary, indicating that they may be effective test cases~\cite{ma2018deepmutation,lu2024mutation}. This motivates us to propose the Demonstration Shuffle mutator that randomly re-orders all demonstrations in the prompt: $(x_i,y_i)\gets (x_{\sigma(i)}, y_{\sigma(i)})$, where $\{\sigma(1),\sigma(2),\cdots,\sigma(k)\}$ is a random permutation of $[1...k]$.

    \item \textbf{Out-of-distribution Demonstrations (OD)}. Similar to the proposed OOD Label mutator, we also consider another form of OOD mutator that introduces a self-consistent OOD demonstration $(x',y')$ from a different dataset, which may also distract the model from the target task: $(x_i, y_i)\gets (x', y'), i\sim Uniform([1...k])$.

    \item \textbf{Demonstration Repetition (DR)}. Finally, we consider the demonstration repetition mutator. The training data repetition mutator was suggested for deep learning with the idea that the same data point might be gathered repeatedly from similar sources~\cite{ma2018deepmutation}. In the case of ICL prompts, repetition or very similar prompts might be seen as unnecessary. As a result, we propose the Demonstration Repetition mutator that incorporates repeated demonstrations into the prompt: $(x_{i+j}, y_{i+j})\gets (x_i, y_i), i\sim Uniform([1...k]), j=1,2,...,N$ where $N$ is the times of repetition. 
\end{itemize}

We present the implementation details of each mutator in experiments in Section~\ref{sec:experiment}. Based on these mutators, we further devise the mutation scores in the testing framework in the following.

\subsection{Mutation Scores}

We first consider the standard mutation score in the context of mutation testing, which is defined as the ratio of mutators killed by (\textit{i.e.} misclassify any case in) the test set. Based on the notations presented in Section~\ref{sec:pre}, this metric can be formulated as:

\begin{definition}[Standard Mutation Score] The standard mutation score $MS_S$ is defined by
\begin{equation}
MS_{S}(M, O, T) = \frac{\#\{o_i|\exists j,\ M'_i(X_j) \ne Y_j\}}{\#O},
\end{equation}
\end{definition}

where $\#S$ denotes the cardinality of set $S$. Please note that in this section we abuse the notation $T$ to denote the test samples that are correctly classified by $M$. Apart from the standard metric, we are also interested in the test set's ability to identify different types of defaults. As outlined in the previous section, the ICL system may have various potential defects. Hence, a high-quality test set should be able to detect a variety of vulnerabilities, measured by the average number of mutator groups killed by the test cases.
Motivated by this notion, we propose a \textbf{Group-wise mutation score} as follows:

\begin{definition}[Group-wise Mutation Score]
Suppose that the mutation operators can be divided into $K$ groups $O=\{O_1,O_2,\cdots, O_K\}$. The group-wise mutation score $MS_G$  is defined by

\begin{equation}
    MS_{G}(M, O, T) = \frac{
    \sum\limits_{i=1}^{\# T} 
    \sum\limits_{j=1}^K 
    \mathbb I(\exists o_l\in O_j, M'_l(X_i)\ne Y_i)
    }{\# T\times K},
\end{equation}
\end{definition}

where $\mathbb I(\cdot)$ is the indicator function. Intuitively, $MS_G$ measures how many groups of mutators can be killed on average, \textit{i.e.}      $\sum\limits_{j=1}^K 
    \mathbb I(\exists o_l\in O_j, M'_l(X_I)\ne Y_i)$. We divide this by $K$ for normalization. This metric underscores the diversity among different mutator groups. This metric is useful for preventing inflation of mutation scores when a test case can only kill mutators from a few groups. In practice, we consider all mutators that are generated from the same operator in the previous section as one group, thus we generally have 6 mutator groups in this testing framework.

%% file: 4_experiments.tex
\section{Experiments}
\label{sec:experiment}

In this section, we conduct evaluations across diverse datasets and LLMs to evaluate and comprehend our MILE framework. We start by elaborating the experiment set-ups, and then showcasing the effectiveness of MILE on measuring dataset quality. Finally, we analyze and compare the mutators for a better understanding of them.

\subsection{Experiment Set-up}

\subsubsection{Datasets.} Following common practice in ICL research~\cite{zhang2024batch}, we consider 5 popular datasets: 

\textbf{(1) SST-2}~\cite{socher2013recursive} (Stanford Sentiment Treebank) is a binary single-sentence classification dataset that is used for sentiment analysis.

\textbf{(2) AGnews}~\cite{zhang2015character} (AG's News Topic Classification Dataset) is a collection of news articles categorized into four different classes: World, Sports, Business, and Sci/Tech.

\textbf{(3) RTE}~\cite{dagan2005pascal} (Recognizing Textual Entailment) contains pairs of sentences where the goal is to determine if the second sentence logically follows from the first.

\textbf{(4) MRPC}~\cite{dolan2005automatically} (Microsoft Research Paraphrase Corpus) is a dataset for text pair classification on whether two sentences are semantically equivalent or not.

\textbf{(5) QNLI}~\cite{wang2018glue} (Question Answering Natural Language Inference) is a dataset for question answering through natural language inference with the task of determining if the answer is supported or contradicted. 

The system prompts and input-label pair formats for these tasks are summarized in Table~\ref{tab:datasets}.

\begin{table}[h]
    \centering
    \caption{System prompts and input-label pair formats for the tasks we used in the experiments.}
    \tabcolsep=10pt
    \renewcommand\arraystretch{1.2}
    \begin{tabular}{l|p{0.4\textwidth}|p{0.3\textwidth}}
    \toprule
        Dataset & System Prompt & Content \\
        \midrule
        SST2~\cite{socher2013recursive} & The following are multiple film reviews with answers (← or →). & Review, Answer\\
        \midrule
        AGnews~\cite{zhang2015character} & Classify the news articles into the categories of 1, 2, 3, or 4. & Title, Description, Answer\\
        \midrule
        RTE~\cite{dagan2005pascal} & Determine whether the hypotheses made based on the premises below are ↑ or ↓. & Premise, Hypothesis, Answer \\
        \midrule 
        MRPC~\cite{dolan2005automatically} & Assess if each pair reflects a semantic equivalence relationship. Use ← or → to indicate the answers.  & Sentence 1, Sentence 2, Answer \\
        \midrule
        QNLI~\cite{wang2018glue} & Please determine whether the paragraph contains the answer to the corresponding question. Use ↑ or ↓ to indicate the answers. & Question, Paragraph, Answer\\
        \bottomrule
    \end{tabular}
    \label{tab:datasets}
\end{table}

\subsubsection{LLMs for evaluation.} We consider 3 popular open-sourced LLMs for our evaluation: \textbf{(1) Vicuna-7b}~\cite{zheng2024judging}, \textbf{(2) Llama-2-chat-7b}~\cite{touvron2023llama} and \textbf{(3) Falcon-7b-instruct}~\cite{almazrouei2023falcon}, which all achieved notable performance across popular LLM benchmarks~\cite{alpaca_eval,dubois2024length}. 

To ensure that these LLMs are capable of conducting ICL on these datasets, we evaluate the vanilla performance of the 3 models on the 5 benchmark datasets with \textbf{20 shots ICL}, as summarized in Table~\ref{tab:acc}. In most cases, they achieve satisfactory accuracy on the tasks, verifying that their ICL inference is reasonable on these datasets. The 20 examples are randomly sampled from the validation set for each task, with the numbers of demonstrations for all classes kept the same. To ensure a fair comparison, we fix these demonstration sets in all following experiments.

\begin{table}[h]
    \centering
    \tabcolsep=10pt
    \renewcommand\arraystretch{1.2}
    \caption{Accuracy evaluation of the 3 LLMs across 5 datasets with vanilla 20 shots ICL.}
    \begin{tabular}{c|ccccc}
    \toprule
    Model & SST2 & AGnews & RTE & MRPC & QNLI \\
    \midrule
    Vicuna & 92.8\% & 68.0\% & 71.2\% & 35.6\% & 56.4\% \\
    Llama-2 & 93.6\% & 61.2\% & 77.2\% & 68.4\% & 63.2\% \\
    Falcon & 78.4\% & 27.6\% & 47.6\% & 68.4\% & 52.4\% \\
    \bottomrule
    \end{tabular}
    \label{tab:acc}
\end{table}

\subsubsection{Mutant implementation details.}  We provide the details of implementing each mutator: 

\textbf{(1)} Noisy Labels (\textbf{NL}): For each input-label demonstration, we randomly flip the label to another possible class in this task and obtain 20 mutant prompts.

\textbf{(2)} OOD Labels \textbf{(OL)}: For each demonstration, we replace the label with a special token '\&', obtaining 20 mutants.

\textbf{(3)} Blurred Inputs \textbf{(BI)}: For each demonstration, we truncate the input with its first-half prefix, getting 20 mutants.

\textbf{(4)} Demonstration Shuffle \textbf{(DS)}: To keep the number of mutants the same as other operators, we randomly generate 20 permutations of $[1...20]$ and apply these orders to the demonstration set.

\textbf{(5)} OOD Demonstrations \textbf{(OD)}: For each demonstration, we replace it with 1 input-output pair randomly sampled from the WMT~\cite{bojar2014findings} dataset, which is a machine translation task from English to France.

\textbf{(6)} Demonstration Repetition \textbf{(DR)}: For each demonstration, we insert two same demonstrations behind it, obtaining 20 mutants.

Finally, with 20 mutants generated by each mutation operator, we collect 120 mutants in total for each vanilla ICL prompt.

\subsection{Overall Assessment}

\subsubsection{Uniform and Non-uniform datasets.}

Our main evaluation aims to evaluate whether the mutation score can reflect the quality of the test set. Following existing evaluation frameworks~\cite{ma2018deepmutation,hu2029deep}, we simulate the quality of the test set through the aspect of the uniformity of the classes. Specifically, a good dataset consists of samples uniformly sampled from all classes, while a dataset consisting of samples from imbalanced classes is considered of poor quality. 

As such, we first construct a dataset that is uniformly sampled from all classes (abbreviated as \textbf{uni.}), and also construct non-uniformly sampled datasets (abbreviated as \textbf{non.}). Specifically, 50\% samples of the dataset are from one single class (called biased class), and another 50\% samples are uniformly sampled from all classes. To make our evaluation results more robust, we create non-uniformly sampled datasets by enumerating all possible biased classes, and then report the average scores across these datasets. We first set the controlled number $n$ as the half-size of the complete dataset, and control the size test set as $\frac 1 2 n$ in our main evaluation. We also investigate the impact of dataset size on the mutation scores in the following.

\subsubsection{Mutation score comparison.}

Based on the settings presented above, we evaluate the standard mutation score ($MS_S$) and group-wise mutation score ($MS_G$) on all datasets and models, and report them in Table~\ref{tab:mss} and Table~\ref{tab:msg} respectively. 

\begin{table}[t]
    \centering
    \tabcolsep=5pt
    \renewcommand\arraystretch{1.2}
    \caption{Standard Mutation Score $MS_S$ comparison between the uniform sampled dataset (uni.) and non-uniformly sampled (non.) dataset.}
    \begin{tabular}{c|cccccc|cc}
    \toprule
        Model & \multicolumn{2}{c}{Vicuna}
        & \multicolumn{2}{c}{Llama-2}
        & \multicolumn{2}{c|}{Falcon}
        & \multicolumn{2}{c}{Average}\\
        Task & uni. & non. 
        & uni. & non. 
        & uni. & non. 
        & uni. & non. \\
    \midrule
SST2 & 54.2\% & 20.4\% & 53.3\% & 20.4\% & 90.0\% & 44.6\% & \textbf{65.8\%} & 28.5\% \\
AGnews & 78.3\% & 36.9\% & 94.2\% & 69.2\% & 60.8\% & 34.2\% & \textbf{77.8\%} & 46.8\% \\
RTE & 47.5\% & 50.4\% & 94.2\% & 85.8\% & 4.2\% & 4.2\% & \textbf{48.6\%} & 46.8\% \\
MRPC & 69.2\% & 50.8\% & 95.0\% & 73.3\% & 75.0\% & 45.8\% & \textbf{79.7\%} & 56.6\% \\
QNLI & 98.3\% & 60.4\% & 95.8\% & 63.3\% & 3.3\% & 3.3\% & \textbf{65.8\%} & 42.3\% \\
\midrule
Avg & \textbf{69.5\% }& 43.8\% & \textbf{86.5\%} & 62.4\% & \textbf{46.7\%} & 26.4\% & \textbf{67.6\%} & 44.2\% \\
    \bottomrule
    \end{tabular}
    \label{tab:mss}
\end{table}

As shown in Table~\ref{tab:mss}, for all tasks the $MS_S$ score of \textbf{uni.} dataset consistently outperforms \textbf{non.} dataset, with 67.6\% \textit{v.s.} 44.2\% on average, indicating a strong correlation between the dataset quality and the mutation score from MILE. Such a significant gap applies to all 3 models, \textit{e.g.} 69.5\% \textit{v.s.} 43.8\% for the Vicuna model, verifying the university of this correlation among different LLMs. For most of the datasets, this property still holds, like the model-averaged score for SST2 exhibits a gap higher than 30\%. There are also exceptional cases like QNLI and RTE tasks for Falcon, where the score is almost the same. However, when reviewing Table~\ref{tab:acc} we can find that Falcon performs poorly on them (near random guess), thus these outliers do not affect our claims.

\begin{table}[t]
    \centering
    \tabcolsep=5pt
    \renewcommand\arraystretch{1.2}
    \caption{Group-wise Mutation Score $MS_G$ comparison between the uniform sampled dataset (uni.) and non-uniformly sampled (non.) dataset.}
    \begin{tabular}{c|cccccc|cc}
    \toprule
        Model & \multicolumn{2}{c}{Vicuna}
        & \multicolumn{2}{c}{Llama-2}
        & \multicolumn{2}{c|}{Falcon}
        & \multicolumn{2}{c}{Average}\\
        Task & uni. & non. 
        & uni. & non. 
        & uni. & non. 
        & uni. & non. \\
    \midrule

SST2 & 13.7\% & 8.7\% & 18.0\% & 10.2\% & 51.0\% & 15.5\% & \textbf{27.6\%} & 11.5\% \\
AGnews & 36.1\% & 10.2\% & 50.3\% & 12.7\% & 53.8\% & 3.8\% & \textbf{46.7\%} & 8.9\% \\
RTE & 10.7\% & 7.3\% & 15.3\% & 12.0\% & 16.7\% & 18.0\% & \textbf{14.2\%} & 12.4\% \\
MRPC & 24.1\% & 13.2\% & 47.7\% & 27.0\% & 70.3\% & 20.8\% & \textbf{47.4\%} & 20.3\% \\
QNLI & 42.7\% & 27.2\% & 34.7\% & 18.3\% & 5.7\% & 9.7\% & \textbf{27.7\%} & 18.4\% \\
\midrule
Avg & \textbf{25.5\%} & 13.3\% & \textbf{33.2\%} & 16.0\% & \textbf{39.5\%} & 13.6\% & \textbf{32.7\%} & 14.3\% \\
    \bottomrule
    \end{tabular}
    \label{tab:msg}
\end{table}

Further, from the group-wise mutation scores in Table~\ref{tab:msg}, we can still observe a strong gap between the scores of \textbf{uni.} and \textbf{non.} datasets, with an averaged score of 32.7\% for \textbf{uni.} to 14.3\% for \textbf{non.} datasets. As a metric with considerations of mutant diversity, the $MS_G$ score also aligns with the superiority of \textbf{uni.} over \textbf{non.} datasets in terms of the comprehensiveness of ICL evaluation. Moreover, the score itself also has an explicit semantic that indicates how many groups of mutants can be detected by each test case on average. For example, since Vicuna achieves 36\% on uni. dataset in the AGnews task, we know that each sample in Vicuna can cover $36\%\times 6\approx 2$ groups of mutants on average.

\subsubsection{Varying dataset size.}

\begin{figure}[h]
    \centering
    \begin{tabular}{ccc}
        \includegraphics[width=0.32\textwidth]{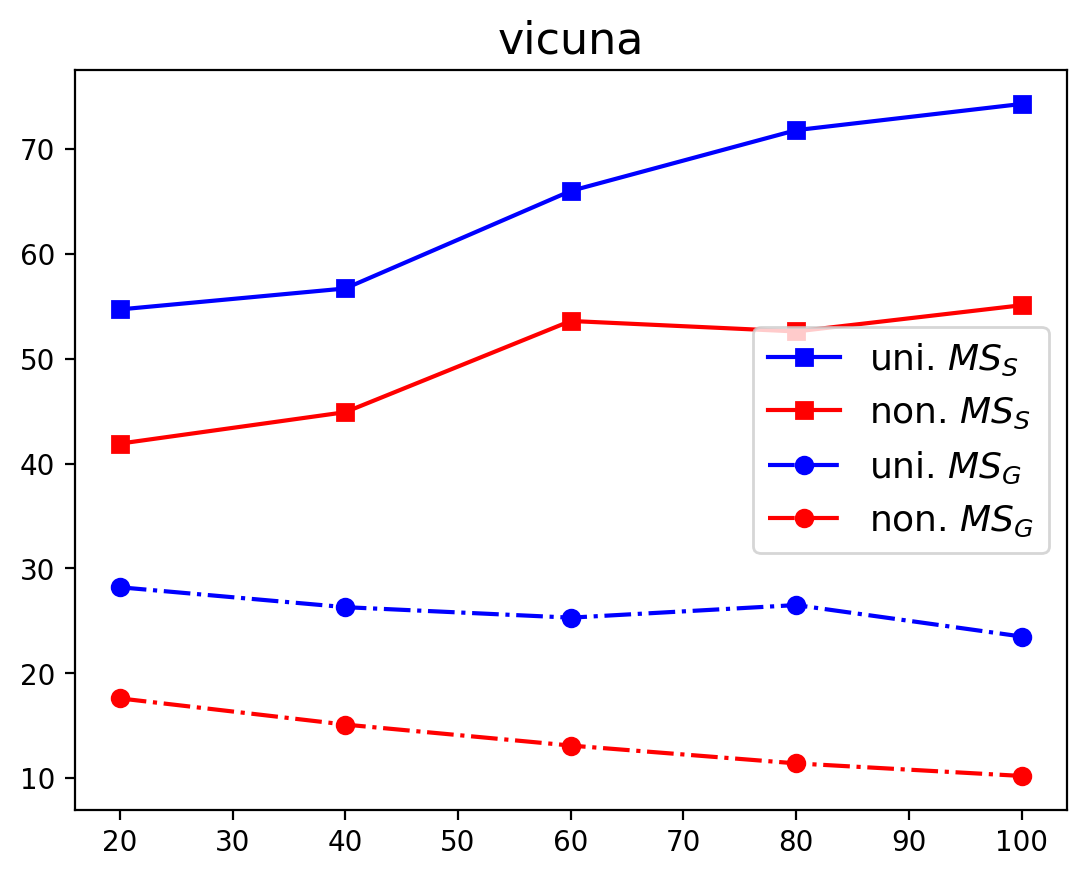}
         & 
         \includegraphics[width=0.32\textwidth]{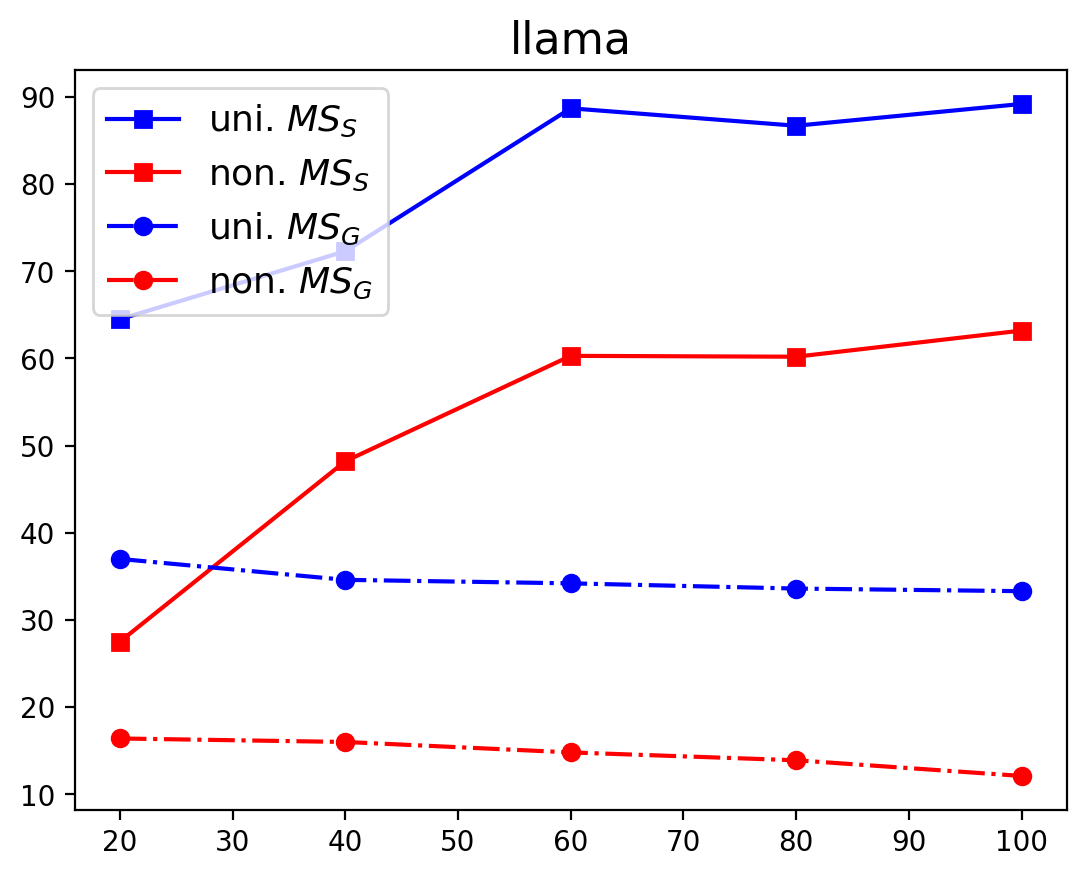}
         &
         \includegraphics[width=0.32\textwidth]{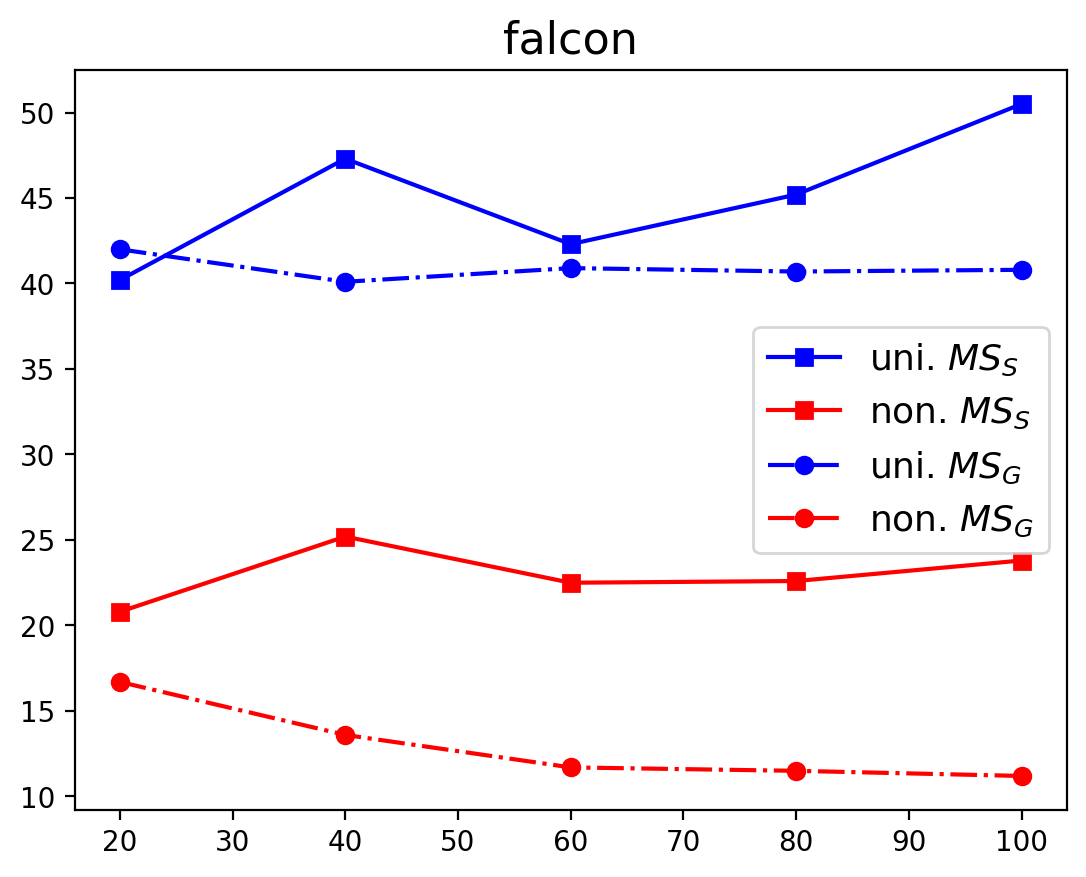} \\
         (a) Vicuna
         &
         (b) Llama-2
         & 
         (c) Falcon
    \end{tabular}
    \caption{Comparing mutation scores with different dataset sizes. Each figure represents the scores averaged over 5 datasets for a model. The X-axis denotes the ratio of the set size to $n$, and the Y-axis denotes the score (\%). The blue lines represent the \textbf{uni.} dataset and red lines represent the \textbf{non.} datasets. The solid line and dotted line denote $MS_S$ and $MS_G$, respectively.}
    \label{fig:size}
\end{figure}

We also conduct an analysis of the scores by varying the size of the test set. To this end, we sampled multiple test sets with sizes $[20\%,40\%, 60\%, 80\%, 100\%]\times n$. The results (averaged over 5 datasets) are summarized by the models in Figure~\ref{fig:size}. For all models, the score superiority of the \textbf{uni.} datasets (blue lines) over \textbf{non.} datasets (red lines) are consistent among different set sizes, further confirming the strong correlation between the scores calculated by MILE. Moreover, an interesting observation is that $MS_S$ gradually increases as the test set becomes larger, since intuitively a larger dataset can cover more mutants. However, the $MS_G$ does not necessarily increase since it is averaged on instance-wise.

\subsection{Mutator Analysis}

In this experiment, we take a closer look at the sensitivity of the ICL model against each mutant group. This analysis aims to better understand the characteristics of each mutation operator, which is beneficial to selecting and allocating mutators for new LLMs or tasks when applying MILE.

Recall that in Section~\ref{sec:framework} we propose the group-wise mutation score as the the average number of mutator groups killed by the test cases. Now, we use a refined metric to analyze the effectiveness of each mutation operator and the corresponding mutants. We first extend the definition of $MS_G$ to the individual cases of a single mutant group $O_j$:

\begin{definition}[Individual Group-wise Mutation Score] 
For a single mutant group $O_j\subset O$, its individual group-wise mutation score is defined as
\begin{equation}
    MS_G(M,O_j,T) = \frac 1 K\sum\limits_{j=1}^K 
    \mathbb I(\exists o_l\in O_j, M'_l(X_i)\ne Y_i).
\end{equation}
\end{definition}

Note that we can rewrite $MS_G(M,O,T) = \frac{1}{\#T}\sum_{j=1}^{\#T} MS_G(M,O_j,T)$. For mutant group $O_j$, the individual $MS_G(M,O_j,T)$ is the proportion of test cases that can kill anyone of the mutants, indicating the sensitivity of the ICL prompt against the mutation operator.

We summarize the individual $MS_G$ scores in Table~\ref{tab:ind} with the scores averaged over 5 datasets. From all models, we can see that the NL (Noisy Label) and (DS) (Demonstration Shuffle) mutators exhibit significantly higher scores than other mutators, which aligns well with the fact that ICL prompts are quite sensitive to label noises~\cite{cheng2024exploring,gao2024noise} and demonstration orders~\cite{lu2021fantastically,fang2024rethinking}. Besides, the OL (OOD Label) mutator achieves a higher score than the other 3 mutators, including the OD (OOD Demonstration) mutator, confirming the sensitivity of the ICL prompts against label perturbation. 

\begin{figure}[h]
    \centering
    \begin{tabular}{ccc}
    \includegraphics[width=0.32\linewidth]{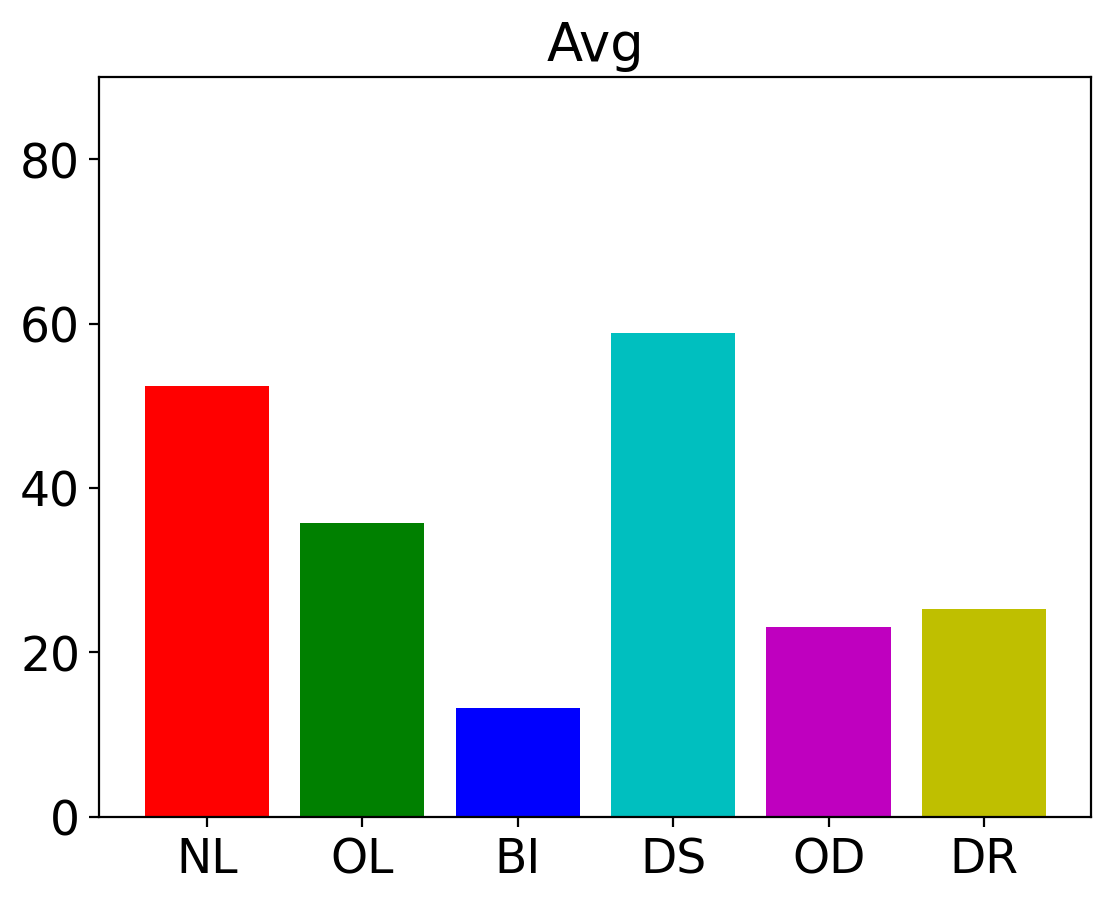} &
    \includegraphics[width=0.32\linewidth]{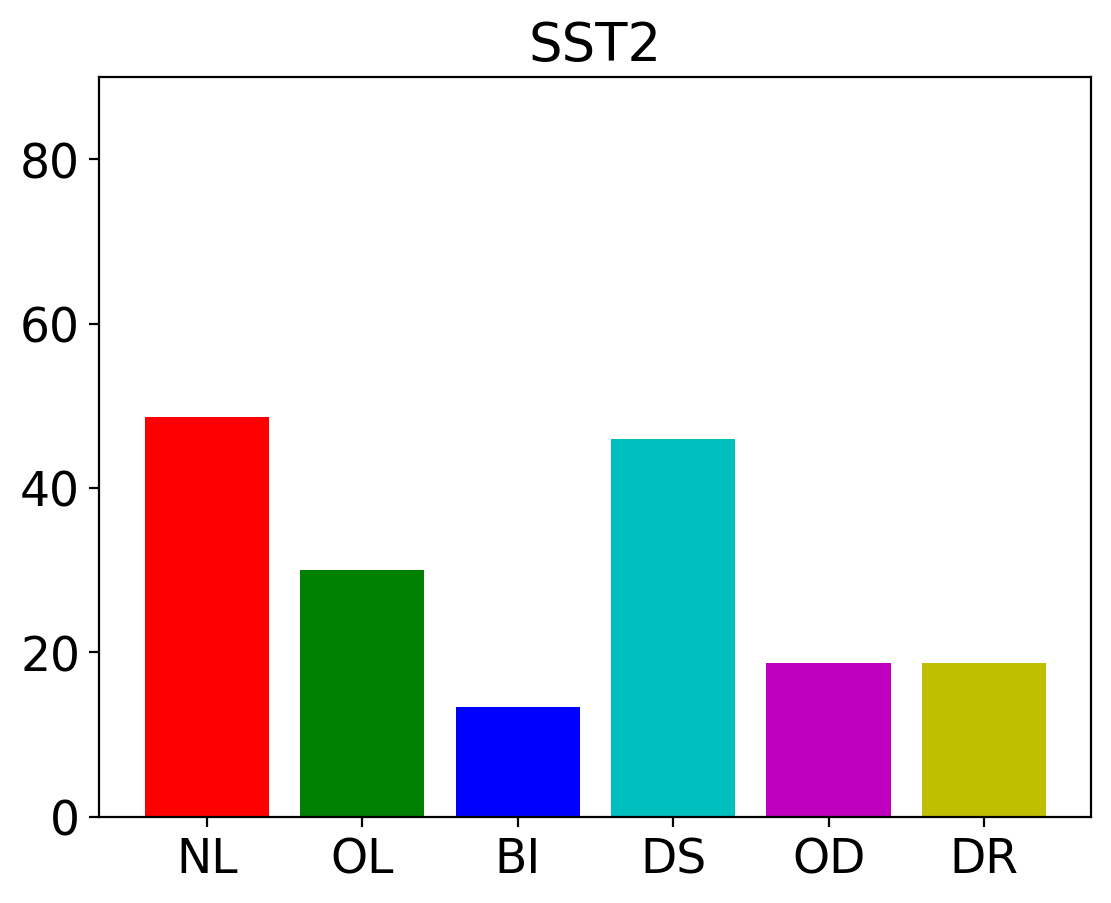} & 
    \includegraphics[width=0.32\linewidth]{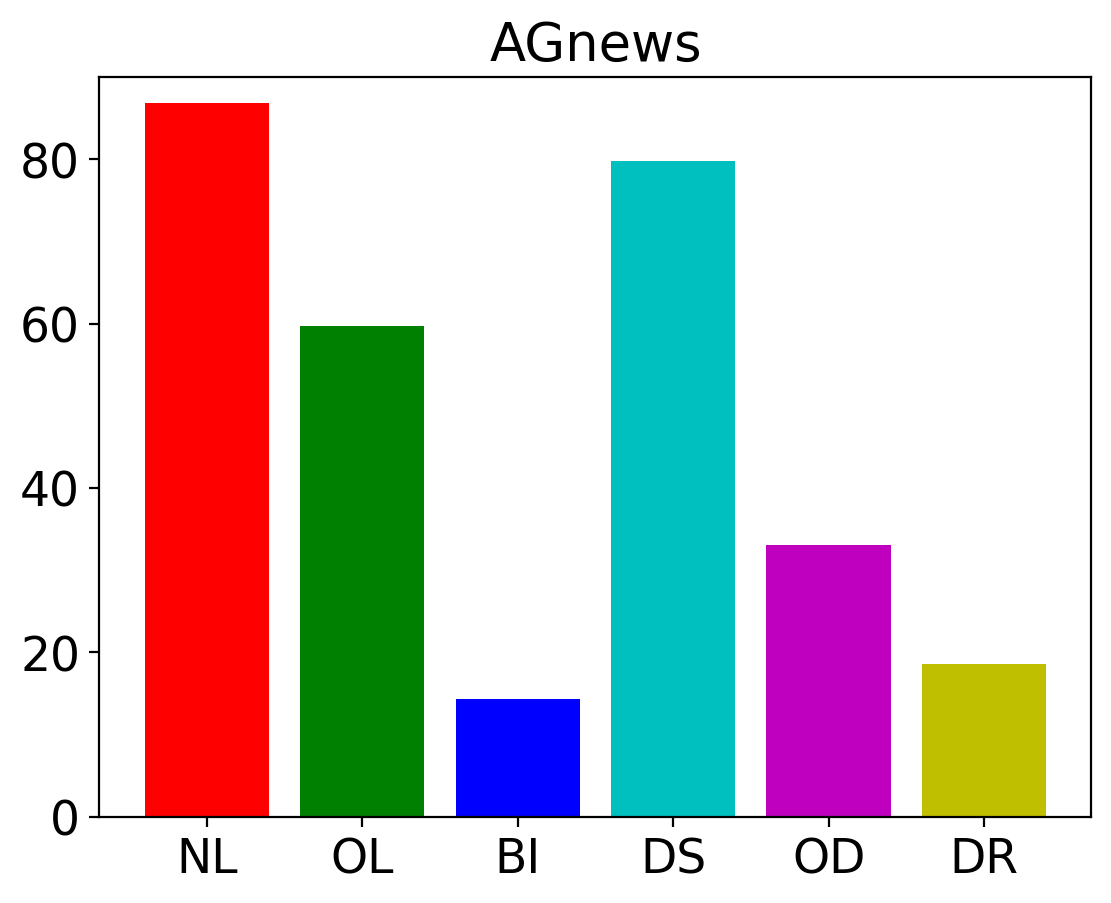} 
     \\
    \vspace{10pt}
    (a) Average & (b) SST-2 & (c) AGnews\\
    \includegraphics[width=0.32\linewidth]{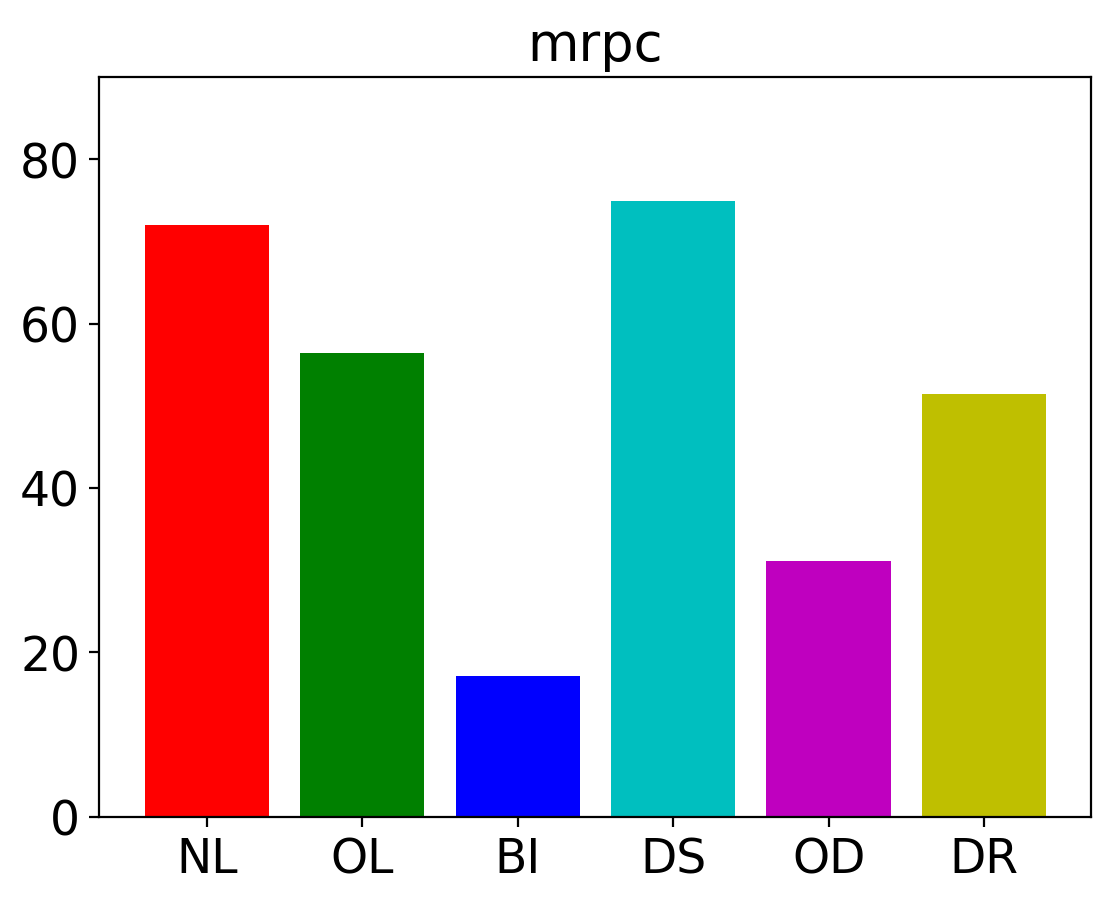}
    & 
    \includegraphics[width=0.32\linewidth]{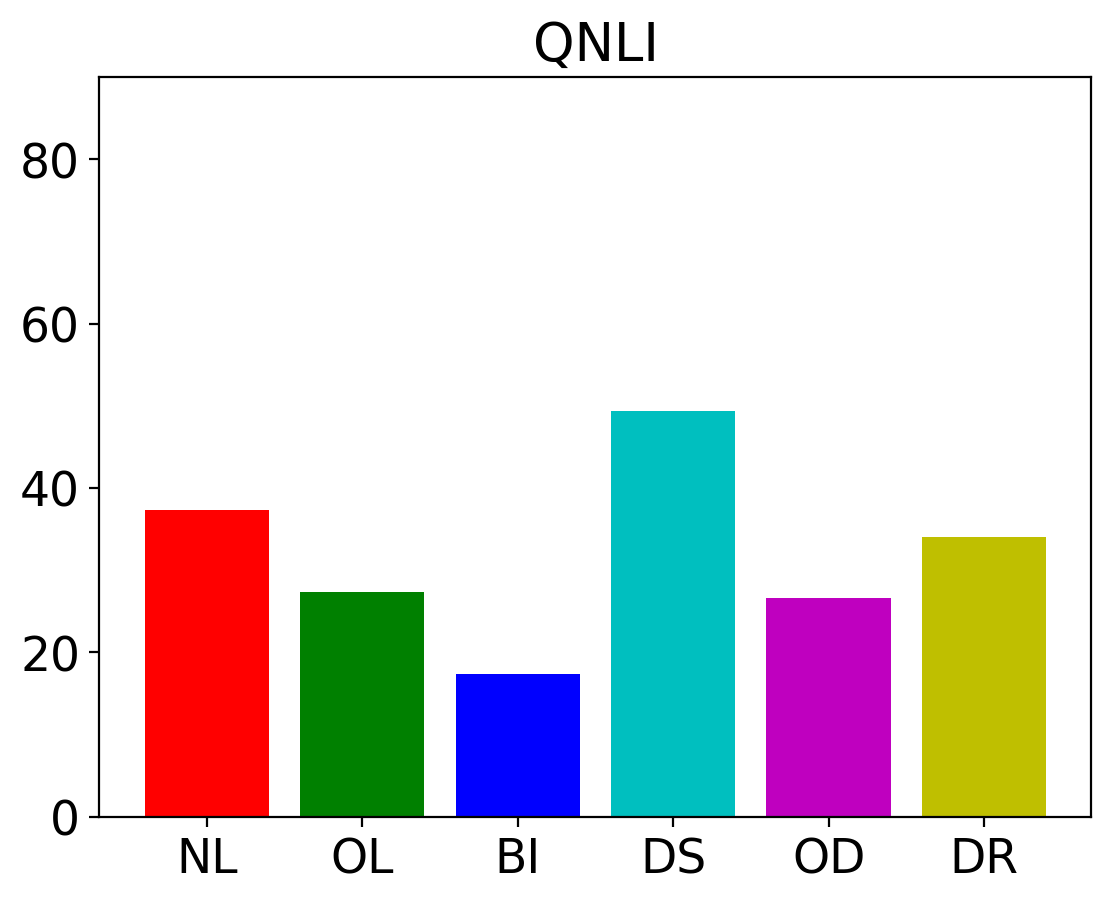}
    &
    \includegraphics[width=0.32\linewidth]{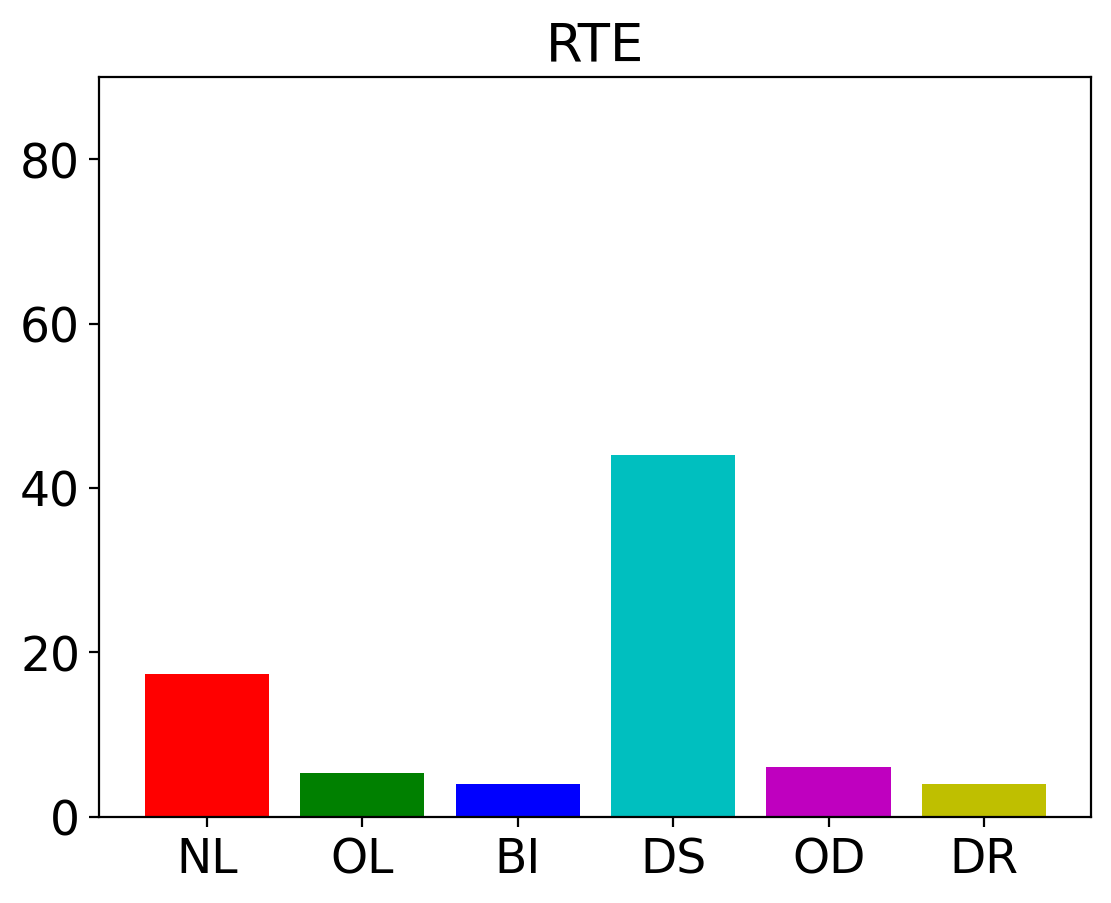}\\
    (d) MRPC & (e) QNLI & (f) RTE
    \end{tabular}
    \caption{Individual $MS_G$ comparison for each mutant group on different datasets. The scores are averaged over 3 models.}
    \label{fig:msg}
\end{figure}

When analyzing this property across different datasets, we can observe the strong transferability of the ranks among the mutators in Figure~\ref{fig:msg}, where the scores are averaged over 3 models. For example, the NL and DS mutators consistently have higher scores than other mutators, verifying the model sensitivity against them across different tasks.

\begin{table}[t]
    \centering
    \tabcolsep=10pt
    \renewcommand\arraystretch{1.2}
    \caption{Individual $MS_G$ comparison for each mutant group. NL: noisy label; OD: OOD label; BI:  blurred input; DS: demo shuffle; OD: OOD demo; DR: demo repetition.}
    \begin{tabular}{c|ccc|ccc}
    \toprule
       Mutator & \multicolumn{3}{c|}{Demonstration-level} & \multicolumn{3}{c}{Prompt-level} \\
       Model & NL & OL &  BI & DS & OD & DR \\
       \midrule
Vicuna & 45.5\% & 23.1\% & 11.6\% & 36.6\% & 23.2\% & 12.9\%   \\
Llama-2 & 50.2\% & 28.4\% & 15.6\% & 70.7\% & 31.0\% & 23.6\%   \\
Falcon & 61.6\% & 55.8\% & 12.5\% & 69.2\% & 15.1\% & 39.5\%   \\
\midrule
Avg. &
52.4\% & 35.8\% & 13.2\% & 58.8\% & 23.1\% & 25.3\%  \\
    \bottomrule
    \end{tabular}
    \label{tab:ind}
\end{table}

In summary, as the mutator sensitivity can be transferred among different models and tasks, we can create a set of mutations for new models and tasks based on specific testing and test set selection needs. For instance, if there's a limited dataset size budget, using more sensitive mutators would help in selecting datasets that can effectively identify faults. On the other hand, using more moderate mutators may be beneficial in designing large-scale datasets to find more nuanced faults.

\subsection{Threats to Validity}
In this paper, we acknowledge the following threats to validity and explain our solutions to them. First, the selection of the LLMs and datasets can be a threat to validity. In our experiments, we have evaluated MILE across 3 LLMs and 5 datasets. Due to computational resource limitations, the models are limited to 7b size, thus selecting a larger model or closed-source model is a potential threat to validity. Besides, the random sampling of the uniform or non-uniform class datasets is also a threat to validity. To deal with this concern, we fixed random seeds in our experiments to ensure reproducibility. Moreover, it is also possible that the model is significantly sensitive or insensitive against some particular biased class during non-uniform sampling. In our experiment, we enumerated all possible biased classes and averaged all scores over these non-uniform datasets to address this issue. Finally, there are also exceptional cases that the score comparison between the two datasets does not align with our overall observation, but when revisiting the vanilla accuracy of the models in these datasets we can find that the model is not capable of conducting reasonable in-context inference on these tasks, and thus would not affect any of our claims.

Overall, we can wrap up the experiment part with the conclusion that the scores from MILE indeed have strong correlations to the test dataset quality, justifying their effectiveness as a metric for dataset quality evaluation. Moreover, we suggest that the mutator sensitivity could be used to generate mutants in new settings.

%% file: 5_related.tex
\section{Related Work}
\label{sec:related}

\subsection{Robustness and Evaluation of In-Context Learning}

Discovered from the GPT-3 model~\cite{brown2020language}, the intriguing ICL ability of LLMs has attracted widespread interest in understanding~\cite{xie2021explanation,min2022rethinking,dai2023can,wang2023large}, utilizing~\cite{wei2022chain,wei2023jailbreak,pawelczyk2023context,wang2024theoretical}, improving~\cite{wei2021finetuned,zhang2024batch,zhao2021calibrate,fang2024rethinking} this learning paradigm. However, though having been studied by a series of works~\cite{fang2024rethinking}, the robustness issue of the in-context demonstrations is still an unaddressed problem. The ICL performance is very sensitive to the selection and order of demonstrations~\cite{lu2021fantastically}, as well as the noise in the labels~\cite{cheng2024exploring,gao2024noise}, both posing safety concerns in their real-world applications. To select better demonstration sets, Zhao et al. attribute the sensitivity to the bias of language models toward predicting certain answers, and propose to fit calibration parameters that cause the prediction to be uniform across classes~\cite{zhao2021calibrate}.
Wang et al. propose to select in-context demonstrations through the Bayesian lens that regard the LLMs as latent variable models~\cite{wang2023large}. There are also other works that attempt to design intrinsically robust ICL against demonstration ordering like Zhang et al. propose BatchICL~\cite{zhang2024batch}, an order-agnostic ICL inference algorithm, and Fang et al. propose InvICL~\cite{fang2024rethinking}, which identifies two crucial factors in the design of ICL including information non-leakage and context interdependence to achieve invariance in ICL.

Apart from focusing on the mechanism of ICL, few works have been dedicated to designing the evaluation specialized for ICL, and most of the existing works still solely conduct ICL evaluation with general LLM benchmarks like Alpaca Eval~\cite{alpaca_eval}, or purely based on conventional natural language processing datasets like SST2. Recently, Chen et al. propose ICLEval~\cite{chen2024icleval}, the first benchmark particularly designed for ICL evaluation with two key sub-abilities of LLMs, including exact copying and rule learning. Besides, the evaluation designed for the quality of test cases for ICL remains unexplored.

\subsection{Mutation Testing for Machine Learning Systems}

In recent years, leveraging mutation testing in machine learning (ML) testing has become a popular research topic~\cite{ma2018deepmutation,zhang2020machine,huang2020survey}. The testing procedure typically consists of 2 stages, including mutating the ML system through different aspects to simulate potential faults within the system, and then evaluating the dataset on the original model and the mutant models to characterize the quality of the dataset or the system. As a pioneering study, Ma et al. propose the DeepMutation~\cite{ma2018deepmutation}, which proposes various mutators for deep neural networks from source-level (training data and model architecture) to model-level (parameters and architecture after training). Then, under controlled experiments, they show that the mutation score is able to reflect the dataset quality for ML 
 systems. Concurrently, Shen et al. propose Munn~\cite{shen2018munn}, including five mutation operators designed with the characteristics of neural networks and investigations on how mutation affects neural networks and how neural depth affects mutation analysis. Subsequently, Humbatova et al. propose DeepCrime~\cite{humbatova2021deepcrime}, which defines 35 deep learning mutation operators and conducts empirical studies about real faults in deep learning systems. 

Going beyond conventional deep learning systems, there are also other works dedicated to applying mutation testing techniques in other learning paradigms and scenarios. Hu et al. propose DeepMutation\texttt{++}~\cite{hu2029deep}, extending the DeepMutation framework to both feed-forward and stateful recurrent neural networks. Lu et al. propose MTUL~\cite{lu2022towards}, 
a mutation testing framework for unsupervised learning systems. Besides, Wang et al.~\cite{wang2019adversarial} propose to leverage mutation testing for adversarial example detection during inference, based on the intuition that adversarial samples are more sensitive against model mutations. Similarly, Zhang et al. propose to apply mutation testing to detect jailbreaking attacks against LLMs~\cite{zhang2023mutation}. On the other position, Yu et al. propose GPUFuzzer~\cite{yu2023gptfuzzer}, leveraging mutation techniques to craft jailbreaking prompts for LLMs. However, although preliminary work has been done on introducing mutation testing for LLMs, the use of mutation testing for ICL systems has not been explored.

%% file: 6_conclusion.tex
\section{Conclusion}
\label{sec:conclusion}

In this paper, we propose MILE, a mutation testing framework of in-context learning (ICL) systems, aiming to evaluate the test suite quality for ICL models. For mutation operators, we consider demonstration-level and prompt-level ones, specialized for ICL prompts. Besides the standard mutation score, we also propose a group-wise mutation score to better understand the model sensitivity against inter-group mutants. With comprehensive experiments across popular LLMs and datasets, we demonstrate the strong correlation between the test set quality and mutation score calculated by MILE, showcasing the effectiveness of using MILE to evaluate the test suite quality. We further investigate the model sensitivity against different kinds of mutants and provide suggestions for designing mutators when applying MILE for different testing goals. Overall, our work provides a new technique for evaluating and improving ICL systems.

%% file: main.bbl
\begin{thebibliography}{10}
\providecommand{\url}[1]{\texttt{#1}}
\providecommand{\urlprefix}{URL }
\providecommand{\doi}[1]{https://doi.org/#1}

\bibitem{achiam2023gpt}
Achiam, J., Adler, S., Agarwal, S., Ahmad, L., Akkaya, I., Aleman, F.L., Almeida, D., Altenschmidt, J., Altman, S., Anadkat, S., et~al.: Gpt-4 technical report. arXiv preprint arXiv:2303.08774  (2023)

\bibitem{agarwal2024many}
Agarwal, R., Singh, A., Zhang, L.M., Bohnet, B., Chan, S., Anand, A., Abbas, Z., Nova, A., Co-Reyes, J.D., Chu, E., et~al.: Many-shot in-context learning. arXiv preprint arXiv:2404.11018  (2024)

\bibitem{agrawal2022context}
Agrawal, S., Zhou, C., Lewis, M., Zettlemoyer, L., Ghazvininejad, M.: In-context examples selection for machine translation. arXiv preprint arXiv:2212.02437  (2022)

\bibitem{almazrouei2023falcon}
Almazrouei, E., Alobeidli, H., Alshamsi, A., Cappelli, A., Cojocaru, R., Debbah, M., Goffinet, {\'E}., Hesslow, D., Launay, J., Malartic, Q., et~al.: The falcon series of open language models. arXiv preprint arXiv:2311.16867  (2023)

\bibitem{anil2024many}
Anil, C., Durmus, E., Sharma, M., Benton, J., Kundu, S., Batson, J., Rimsky, N., Tong, M., Mu, J., Ford, D., et~al.: Many-shot jailbreaking. Anthropic  (2024)

\bibitem{bojar2014findings}
Bojar, O., Buck, C., Federmann, C., Haddow, B., Koehn, P., Leveling, J., Monz, C., Pecina, P., Post, M., Saint-Amand, H., et~al.: Findings of the 2014 workshop on statistical machine translation. In: Proceedings of the ninth workshop on statistical machine translation (2014)

\bibitem{brown2020language}
Brown, T.B.: Language models are few-shot learners. arXiv preprint arXiv:2005.14165  (2020)

\bibitem{chen2024icleval}
Chen, W., Lin, Y., Zhou, Z., Huang, H., Jia, Y., Cao, Z., Wen, J.R.: Icleval: Evaluating in-context learning ability of large language models. arXiv preprint arXiv:2406.14955  (2024)

\bibitem{cheng2024exploring}
Cheng, C., Yu, X., Wen, H., Sun, J., Yue, G., Zhang, Y., Wei, Z.: Exploring the robustness of in-context learning with noisy labels. In: ICLR 2024 Workshop on Reliable and Responsible Foundation Models (2024)

\bibitem{dagan2005pascal}
Dagan, I., Glickman, O., Magnini, B.: The pascal recognising textual entailment challenge. In: Machine learning challenges workshop (2005)

\bibitem{dai2023can}
Dai, D., Sun, Y., Dong, L., Hao, Y., Ma, S., Sui, Z., Wei, F.: Why can gpt learn in-context? language models secretly perform gradient descent as meta-optimizers. In: ACL (2023)

\bibitem{demillo1991constraint}
DeMillo, R.A., Offutt, A.J., et~al.: Constraint-based automatic test data generation. IEEE Transactions on Software Engineering  (1991)

\bibitem{dolan2005automatically}
Dolan, B., Brockett, C.: Automatically constructing a corpus of sentential paraphrases. In: Third international workshop on paraphrasing (IWP2005) (2005)

\bibitem{dong2023survey}
Dong, Q., Li, L., Dai, D., Zheng, C., Wu, Z., Chang, B., Sun, X., Xu, J., Li, L., Sui, Z.: A survey on in-context learning (2023)

\bibitem{dubois2024length}
Dubois, Y., Galambosi, B., Liang, P., Hashimoto, T.B.: Length-controlled alpacaeval: A simple way to debias automatic evaluators. arXiv preprint arXiv:2404.04475  (2024)

\bibitem{fang2024rethinking}
Fang, L., Wang, Y., Gatmiry, K., Fang, L., Wang, Y.: Rethinking invariance in in-context learning. In: ICML 2024 Workshop on Theoretical Foundations of Foundation Models (2024)

\bibitem{gao2024noise}
Gao, H., Zhang, F., Jiang, W., Shu, J., Zheng, F., Wei, H.: On the noise robustness of in-context learning for text generation. arXiv preprint arXiv:2405.17264  (2024)

\bibitem{garg2022can}
Garg, S., Tsipras, D., Liang, P.S., Valiant, G.: What can transformers learn in-context? a case study of simple function classes. NeurIPS  (2022)

\bibitem{ghanbari2019practical}
Ghanbari, A., Benton, S., Zhang, L.: Practical program repair via bytecode mutation. In: ISSTA (2019)

\bibitem{hu2029deep}
Hu, Q., Ma, L., Xie, X., Yu, B., Liu, Y., Zhao, J.: Deepmutation++: A mutation testing framework for deep learning systems. In: ASE (2019)

\bibitem{huang2020survey}
Huang, X., Kroening, D., Ruan, W., Sharp, J., Sun, Y., Thamo, E., Wu, M., Yi, X.: A survey of safety and trustworthiness of deep neural networks: Verification, testing, adversarial attack and defence, and interpretability. Computer Science Review  (2020)

\bibitem{huang2024understanding}
Huang, X., Liu, W., Chen, X., Wang, X., Wang, H., Lian, D., Wang, Y., Tang, R., Chen, E.: Understanding the planning of llm agents: A survey. arXiv preprint arXiv:2402.02716  (2024)

\bibitem{humbatova2021deepcrime}
Humbatova, N., Jahangirova, G., Tonella, P.: Deepcrime: mutation testing of deep learning systems based on real faults. In: ISSTA (2021)

\bibitem{jia2010analysis}
Jia, Y., Harman, M.: An analysis and survey of the development of mutation testing. IEEE transactions on software engineering  \textbf{37}(5) (2010)

\bibitem{just2014defects4j}
Just, R., Jalali, D., Ernst, M.D.: Defects4j: A database of existing faults to enable controlled testing studies for java programs. In: ISSTA (2014)

\bibitem{alpaca_eval}
Li, X., Zhang, T., Dubois, Y., Taori, R., Gulrajani, I., Guestrin, C., Liang, P., Hashimoto, T.B.: Alpacaeval: An automatic evaluator of instruction-following models. \url{https://github.com/tatsu-lab/alpaca_eval}

\bibitem{liu2020energy}
Liu, W., Wang, X., Owens, J., Li, Y.: Energy-based out-of-distribution detection. NeurIPS  (2020)

\bibitem{lu2023emergent}
Lu, S., Bigoulaeva, I., Sachdeva, R., Madabushi, H.T., Gurevych, I.: Are emergent abilities in large language models just in-context learning? arXiv preprint arXiv:2309.01809  (2023)

\bibitem{lu2021fantastically}
Lu, Y., Bartolo, M., Moore, A., Riedel, S., Stenetorp, P.: Fantastically ordered prompts and where to find them: Overcoming few-shot prompt order sensitivity. arXiv preprint arXiv:2104.08786  (2021)

\bibitem{lu2024mutation}
Lu, Y., Shao, K., Zhao, J., Sun, W., Sun, M.: Mutation testing of unsupervised learning systems. Journal of Systems Architecture  \textbf{146},  103050 (2024)

\bibitem{lu2022towards}
Lu, Y., Sun, W., Sun, M.: Towards mutation testing of reinforcement learning systems. Journal of Systems Architecture  \textbf{131},  102701 (2022)

\bibitem{ma2018deepmutation}
Ma, L., Zhang, F., Sun, J., Xue, M., Li, B., Juefei-Xu, F., Xie, C., Li, L., Liu, Y., Zhao, J., et~al.: Deepmutation: Mutation testing of deep learning systems. In: ISSRE (2018)

\bibitem{min2022rethinking}
Min, S., Lyu, X., Holtzman, A., Artetxe, M., Lewis, M., Hajishirzi, H., Zettlemoyer, L.: Rethinking the role of demonstrations: What makes in-context learning work? In: EMNLP (2022)

\bibitem{offutt2001mutation}
Offutt, A.J., Untch, R.H.: Mutation 2000: Uniting the orthogonal. Mutation testing for the new century  (2001)

\bibitem{papadakis2015metallaxis}
Papadakis, M., Le~Traon, Y.: Metallaxis-fl: mutation-based fault localization. Software Testing, Verification and Reliability  \textbf{25} (2015)

\bibitem{pawelczyk2023context}
Pawelczyk, M., Neel, S., Lakkaraju, H.: In-context unlearning: Language models as few shot unlearners. arXiv preprint arXiv:2310.07579  (2023)

\bibitem{qiang2023hijacking}
Qiang, Y., Zhou, X., Zhu, D.: Hijacking large language models via adversarial in-context learning. CoRR  (2023)

\bibitem{ratner2023parallel}
Ratner, N., Levine, Y., Belinkov, Y., Ram, O., Magar, I., Abend, O., Karpas, E., Shashua, A., Leyton-Brown, K., Shoham, Y.: Parallel context windows for large language models. In: ACL (2023)

\bibitem{schaeffer2024emergent}
Schaeffer, R., Miranda, B., Koyejo, S.: Are emergent abilities of large language models a mirage? NeurIPS  (2024)

\bibitem{shen2018munn}
Shen, W., Wan, J., Chen, Z.: Munn: Mutation analysis of neural networks. In: QRS-C (2018)

\bibitem{socher2013recursive}
Socher, R., Perelygin, A., Wu, J., Chuang, J., Manning, C.D., Ng, A.Y., Potts, C.: Recursive deep models for semantic compositionality over a sentiment treebank. In: EMNLP (2013)

\bibitem{touvron2023llama}
Touvron, H., Martin, L., Stone, K., Albert, P., Almahairi, A., Babaei, Y., Bashlykov, N., Batra, S., Bhargava, P., Bhosale, S., et~al.: Llama 2: Open foundation and fine-tuned chat models. arXiv preprint arXiv:2307.09288  (2023)

\bibitem{uesato2018rigorous}
Uesato, J., Kumar, A., Szepesvari, C., Erez, T., Ruderman, A., Anderson, K., Heess, N., Kohli, P., et~al.: Rigorous agent evaluation: An adversarial approach to uncover catastrophic failures. arXiv preprint arXiv:1812.01647  (2018)

\bibitem{wang2018glue}
Wang, A., Singh, A., Michael, J., Hill, F., Levy, O., Bowman, S.R.: Glue: A multi-task benchmark and analysis platform for natural language understanding. arXiv preprint arXiv:1804.07461  (2018)

\bibitem{wang2019adversarial}
Wang, J., Dong, G., Sun, J., Wang, X., Zhang, P.: Adversarial sample detection for deep neural network through model mutation testing. In: ICSE (2019)

\bibitem{wang2023adversarial}
Wang, J., Liu, Z., Park, K.H., Jiang, Z., Zheng, Z., Wu, Z., Chen, M., Xiao, C.: Adversarial demonstration attacks on large language models. arXiv preprint arXiv:2305.14950  (2023)

\bibitem{wangexecutable}
Wang, X., Chen, Y., Yuan, L., Zhang, Y., Li, Y., Peng, H., Ji, H.: Executable code actions elicit better llm agents. In: ICML (2024)

\bibitem{wang2023large}
Wang, X., Zhu, W., Wang, W.Y.: Large language models are implicitly topic models: Explaining and finding good demonstrations for in-context learning. arXiv preprint arXiv:2301.11916  (2023)

\bibitem{wang2024theoretical}
Wang, Y., Wu, Y., Wei, Z., Jegelka, S., Wang, Y.: A theoretical understanding of self-correction through in-context alignment. arXiv preprint arXiv:2405.18634  (2024)

\bibitem{wei2021finetuned}
Wei, J., Bosma, M., Zhao, V.Y., Guu, K., Yu, A.W., Lester, B., Du, N., Dai, A.M., Le, Q.V.: Finetuned language models are zero-shot learners. arXiv preprint arXiv:2109.01652  (2021)

\bibitem{wei2022chain}
Wei, J., Wang, X., Schuurmans, D., Bosma, M., Xia, F., Chi, E., Le, Q.V., Zhou, D., et~al.: Chain-of-thought prompting elicits reasoning in large language models. NeurIPS  (2022)

\bibitem{wei2023jailbreak}
Wei, Z., Wang, Y., Wang, Y.: Jailbreak and guard aligned language models with only few in-context demonstrations. arXiv preprint arXiv:2310.06387  (2023)

\bibitem{wu2023autogen}
Wu, Q., Bansal, G., Zhang, J., Wu, Y., Zhang, S., Zhu, E., Li, B., Jiang, L., Zhang, X., Wang, C.: Autogen: Enabling next-gen llm applications via multi-agent conversation framework. arXiv preprint arXiv:2308.08155  (2023)

\bibitem{xie2021explanation}
Xie, S.M., Raghunathan, A., Liang, P., Ma, T.: An explanation of in-context learning as implicit bayesian inference. arXiv preprint arXiv:2111.02080  (2021)

\bibitem{yang2024harnessing}
Yang, J., Jin, H., Tang, R., Han, X., Feng, Q., Jiang, H., Zhong, S., Yin, B., Hu, X.: Harnessing the power of llms in practice: A survey on chatgpt and beyond. ACM Transactions on Knowledge Discovery from Data  (2024)

\bibitem{yang2024generalized}
Yang, J., Zhou, K., Li, Y., Liu, Z.: Generalized out-of-distribution detection: A survey. International Journal of Computer Vision  (2024)

\bibitem{yu2023gptfuzzer}
Yu, J., Lin, X., Xing, X.: Gptfuzzer: Red teaming large language models with auto-generated jailbreak prompts. arXiv preprint arXiv:2309.10253  (2023)

\bibitem{zhang2020machine}
Zhang, J.M., Harman, M., Ma, L., Liu, Y.: Machine learning testing: Survey, landscapes and horizons. IEEE Transactions on Software Engineering  \textbf{48} (2020)

\bibitem{zhang2024batch}
Zhang, K., Lv, A., Chen, Y., Ha, H., Xu, T., Yan, R.: Batch-icl: Effective, efficient, and order-agnostic in-context learning. In: ACL (2024)

\bibitem{zhang2015character}
Zhang, X., Zhao, J., LeCun, Y.: Character-level convolutional networks for text classification. NeurIPS  (2015)

\bibitem{zhang2023mutation}
Zhang, X., Zhang, C., Li, T., Huang, Y., Jia, X., Xie, X., Liu, Y., Shen, C.: A mutation-based method for multi-modal jailbreaking attack detection. arXiv preprint arXiv:2312.10766  (2023)

\bibitem{zhao2021calibrate}
Zhao, Z., Wallace, E., Feng, S., Klein, D., Singh, S.: Calibrate before use: Improving few-shot performance of language models. In: ICML (2021)

\bibitem{zheng2024judging}
Zheng, L., Chiang, W.L., Sheng, Y., Zhuang, S., Wu, Z., Zhuang, Y., Lin, Z., Li, Z., Li, D., Xing, E., et~al.: Judging llm-as-a-judge with mt-bench and chatbot arena. NeurIPS  (2024)

\end{thebibliography}
